\documentclass[letter,twocolumn]{jpsj3}
\usepackage{txfonts}
\usepackage{amssymb}
\def\Vec#1{\mbox{\boldmath $#1$}}
\def\cH{{\mathcal H}}
\def\cI{{\mathcal I}}
\def\cT{{\mathcal T}}
\def\vS{{\Vec S}}

\title{Ground-State Phase Diagram of the Bond-Alternating $S=2$ Quantum Spin Chain
with the $XXZ$ and On-Site Anisotropies --- Symmetry Protected Topological Phase versus Trivial Phase ---}
\author{Kiyomi Okamoto$^1$\thanks{nnn2411@yahoo.co.jp}, 
   Takashi Tonegawa,$^{2,3}$\thanks{tone0115@vivid.ocn.ne.jp} 
   and T\^oru Sakai$^{4,5}$\thanks{sakai@spring8.or.jp}}
\inst{$^1$School of Arts and Sciences, College of Engineering, Shibaura Institute of Technology, 
      Minuma-ku, Saitama-shi, Saitama, 337-8570, Japan \\
      $^2$Professor Emeritus, Kobe University, Nada-ku, Kobe-shi, Hyogo,  657-8501, Japan \\
      $^3$Department of Physical Science, School of Science, Osaka Prefecture University, Naka-ku, Sakai-shi, 
      Osaka, 599-8531, Japan \\
      $^4$Graduate School of Material Science, University of Hyogo, Kamigori-cho, Ako-gun, Hyogo,
      678-1297, Japan \\
      $^5$National Institutes for Quantum and Radiological Science and Technology, 
      SPring-8, Sayo-cho, Sayo-gun, Hyogo, 679-5148, Japan }
\abst{We investigate the ground-state phase diagram of
the bond-alternating $S=2$ quantum spin chain with the $XXZ$ and on-site anisotropies.
For the on-site anisotropies, in addition to the popular $D_2 \sum_j (S_j^z)^2$ term,
we consider the $D_4 \sum_j (S_j^z)^4$ term.
Mainly we use the exact diagonalization and the level spectroscopy analysis.
We show that the Haldane state, large-$D$ state and the Dimer2 state belong to the same trivial phase,
by finding the existence of adiabatic paths directly connecting these
states without the quantum phase transition.
Similarly, we show that the intermediate-$D$ state and the Dimer1 state
belong to the same symmetry protected topological phase.
}

\begin{document}
\maketitle

\begin{figure}[ht]
       \centerline{
       \includegraphics[height=10pc]{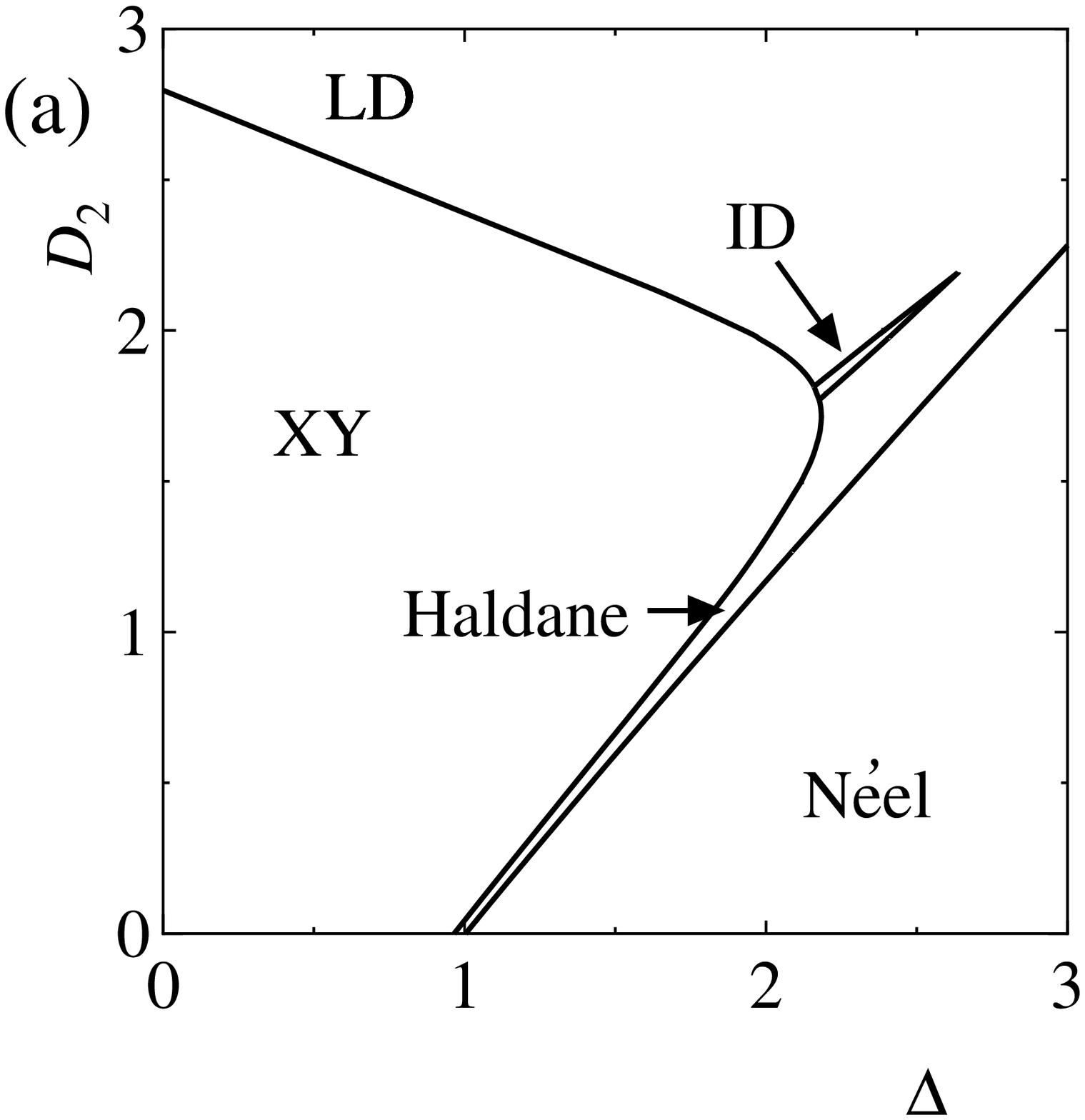}
       \includegraphics[height=10pc]{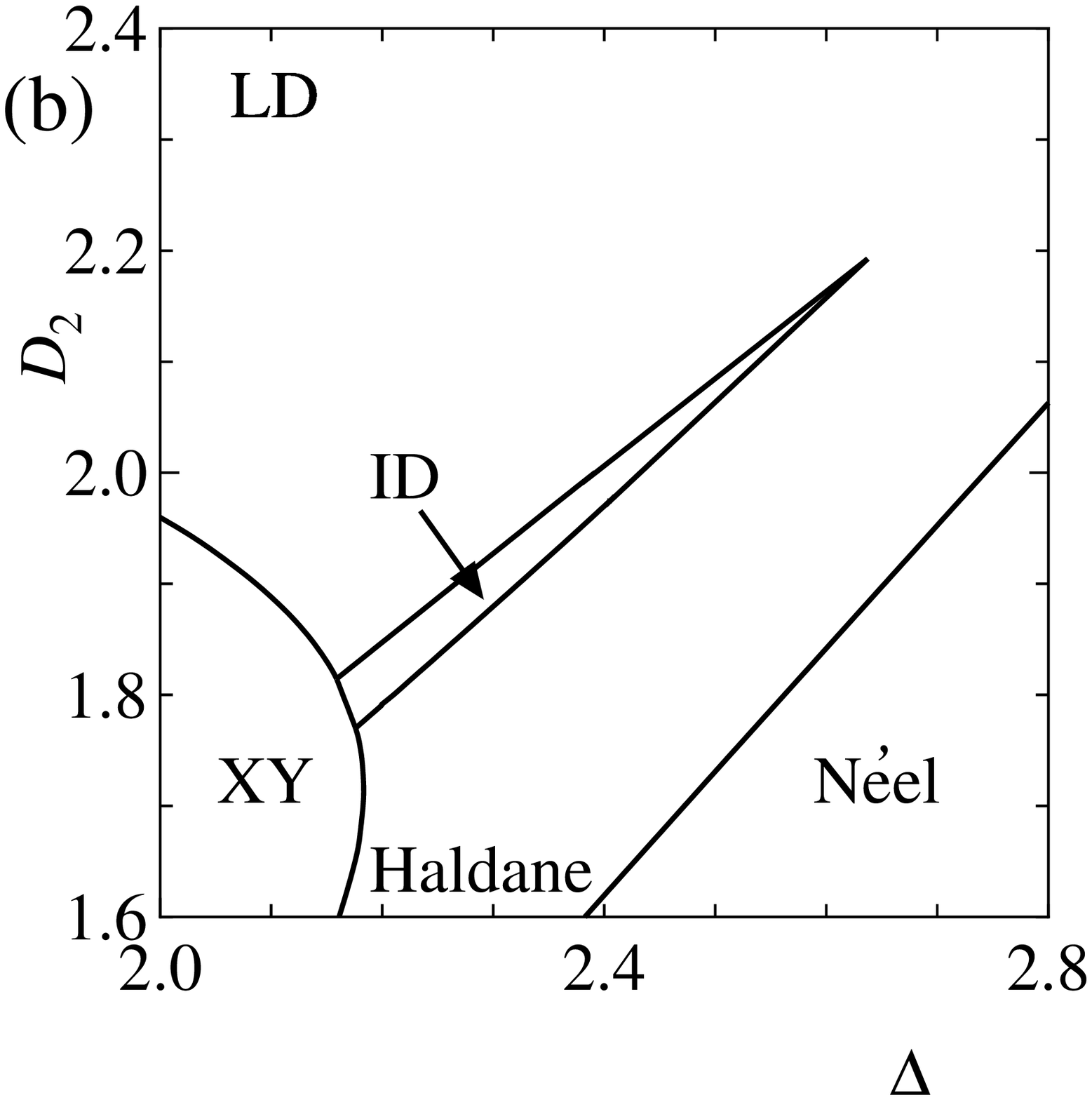}}
       \caption{GS phase diagram\cite{tone,oka1} of the $S=2$ chain described by the Hamiltonian (\ref{eq:Ham-1}).
       (a) is a wide view, while (b) is an enlarged view near the ID region.}
       \label{fig:fig1}
\end{figure}
\begin{figure}[ht]
       \centerline{
       \includegraphics[width=11pc]{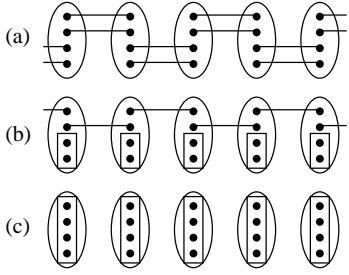}}%
       \caption{\label{fig:vbs-pictures-1}
       Valence bond pictures for (a) the Haldane state, (b) the ID state
       and (c) the LD state.  Big ellipses denote $S=2$ spins and dots
       $S=1/2$ spins.  Solid lines represent valence bonds
       (singlet pairs of two $=1/2$ spins,
       $(1/\sqrt{2})(\uparrow\downarrow-\downarrow\uparrow)$).
       Two $S=1/2$ spins in rectangles are in the
       $(S_{\rm tot},S^z_{\rm tot})=(1,0)$ state
       and similarly four $S=1/2$ spins in rectangles are in the $(S_{\rm tot},S^z_{\rm tot})=(2,0)$ state.}
\end{figure}

In these years, quantum spin chain systems have been attracting increasing attention
because they provide rich physics even when models are rather simple.
Recently we investigated\cite{tone,oka1,oka2,oka3} the $S=2$ quantum spin chain
 with the $XXZ$ and on-site anisotropies
described by
\begin{equation}
    \cH_1
    = \sum_j (S_j^x S_{j+1}^x + S_j^y S_{j+1}^y +\Delta S_j^z S_{j+1}^z)
      + D_2 \sum_j (S_j^z)^2,
    \label{eq:Ham-1}
\end{equation}
where $S_j^\mu$ ($\mu = x,y,z$) represents the $\mu$-component of the $S=2$
operator $\vS_j$ at the $j$th site, and $\Delta$ and $D_2$ are, respectively, the
$XXZ$ anisotropy parameter of the nearest-neighbor interactions and the on-site anisotropy parameter.
Our ground-state (GS) phase diagram\cite{tone,oka1} obtained mainly by the use of the exact diagonalization and the
level spectroscopy (LS) analysis\cite{okamoto-ls,nomura-ls,kitazawa-ls,nomura-kitazawa-ls}
is shown in Fig.\ref{fig:fig1},
where we restrict ourselves to the $\Delta \ge 0$ and $D_2 \ge 0$ case, for simplicity.
There are four phases in this GS phase diagram,
the $XY$ phase, the N\'eel phase, the Haldane/Large-$D$ (H/LD) phase and the intermediate-$D$ (ID) phase.
The valence-bond pictures of the Haldane state, the ID state and the LD state
are depicted in Fig.\ref{fig:vbs-pictures-1}.
The remarkable features of the GS phase diagram shown in Fig.\ref{fig:fig1} are:
(a) there exists the ID phase which was predicted by Oshikawa in 1992 and has
been believed to be absent for about two decades until our finding in 2011; 
(b) the Haldane state and the LD state belong to
the same phase. 
These features are consistent with the discussion by Pollmann {\it et al}.\cite{pollmann2010,pollmann2012}
Namely, they showed the existence of a symmetry-protected topological (SPT) state
if any one of the following three global symmetries is satisfied: 
(i) the dihedral group of $\pi$ rotations about the $x$, $y$, and $z$ axes, 
(ii) the time-reversal symmetry $\vS_j \to -\vS_j$,
and (iii) the space inversion symmetry with respect to a bond.
It is easy to see that the Hamiltonian (\ref{eq:Ham-1}) satisfies (ii) and (iii),
but not (i).
In the GS phase diagram shown in Fig.\ref{fig:fig1},
the ID phase is the SPT phase and the H/LD phase is the trivial phase.
For (b),
Pollmann {\it et al}.\cite{pollmann2012} constructed a one-parameter matrix product state
which interpolates the Haldane and LD states without any
quantum phase transition.

Slightly after our works\cite{tone,oka1,oka2}, 
Tzeng\cite{tzeng} confirmed our results by use of the parity density matrix
renormalization group (DMRG) and the LS analysis.
Kj\"all {\it et al}.\cite{kjall} also studied the Hamiltonian (\ref{eq:Ham-1})
by use of the DMRG based on the matrix product state.
They obtained the same conclusion as ours with respect to (b),
whereas somewhat different one from ours and Tzeng's with respect to (a).
Namely, they stated that the ID phase does not exist on the $\Delta-D_2$ plane
and very small positive $D_4$ (see eq.(\ref{eq:Ham-3})) is necessary to realize the ID state,
although they avoid the definite conclusion.
Nevertheless, we think that the difference
between our and their conclusions for (a) is not a serious problem\cite{oka3}
as will be discussed later.

\begin{figure}[ht]
       \centerline{
       \includegraphics[width=11pc]{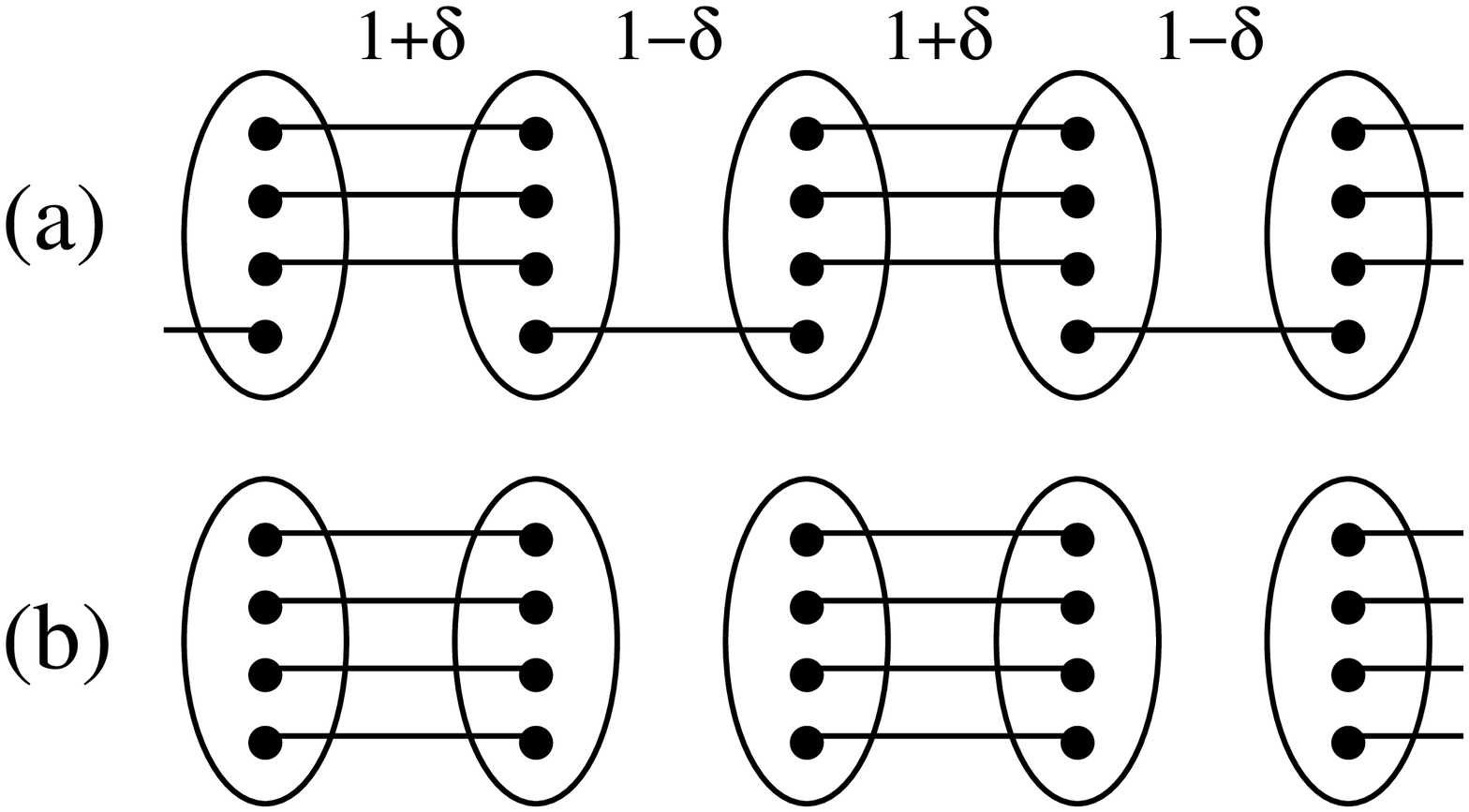}}%
       \caption{\label{fig:vbs-pictures-2}
       Valence bond pictures for (a) the dimer1 (Dim1) state, and (b) the dimer2 (Dim2) state.
       }
\end{figure}

The bond alternating isotropic $S=2$ quantum spin chain
\begin{equation}
    \cH_2
    = \sum_j [1+(-1)^j\delta] \vS_j \cdot \vS_{j+1},
    \label{eq:Ham-2}
\end{equation}
where $\delta$ is the bond alternation parameter,
has been investigated by several authors.\cite{yamanaka,yamamoto,kitazawa,nakamura}
With the increase of $\delta$ from 0 to 1,
the first transition from the Haldane state to the dimer1 (Dim1) state 
occurs at $\delta^{\rm (cr,1)}$,
and after that, the second transition from the Dim1 state to the dimer2 (Dim2) state takes place
at $\delta^{\rm (cr,2)}$.
According to Ref.\citen{nakamura},
these critical values are $\delta^{\rm (cr,1)} = 0.1866 \pm 0.0007$
and $\delta^{\rm (cr,2)} = 0.5500 \pm 0.0001$,
respectively, which are very similar to those of other works.\cite{yamamoto,kitazawa}
The valence bond pictures of the Dim1 and Dim2 states are depicted in Fig.\ref{fig:vbs-pictures-2}.
The Dim1 state is the SPT state, while the Haldane state and the Dim2 state are the trivial states.
We note that the Hamiltonian $\cH_2$ satisfies the conditions (i), (ii) and (iii) 
by Pollmann {\it et al}.

From the standpoint of the SPT state,
it is strongly expected that
the H/LD state and the Dim2 state belong to the same trivial phase,
while the ID state and the Dim1 state to the same SPT phase.
The clearest evidence for the above prediction
is the existence of direct adiabatic paths connecting the
H/LD state and the Dim2 state, and connecting the ID state and the Dim1 state.
To prove this,
we investigate the following Hamiltonian
\begin{eqnarray}
    \cH_3
    &=& \sum_j [1+(-1)^j\delta](S_j^x S_{j+1}^x + S_j^y S_{j+1}^y +\Delta S_j^z S_{j+1}^z) \nonumber \\
    &&  + \,D_2 \sum_j (S_j^z)^2
      + D_4 \sum_j (S_j^z)^4,
    \label{eq:Ham-3}
\end{eqnarray}
which satisfies the conditions (ii) and (iii)  of Pollmann {\it et al}.
We mainly use numerical methods based on the exact diagonalization calculation for finite spin
systems with up to $N = 12$, where $N$ is the number of spins.

\begin{figure}[ht]
       \centerline{
       \includegraphics[width=10pc]{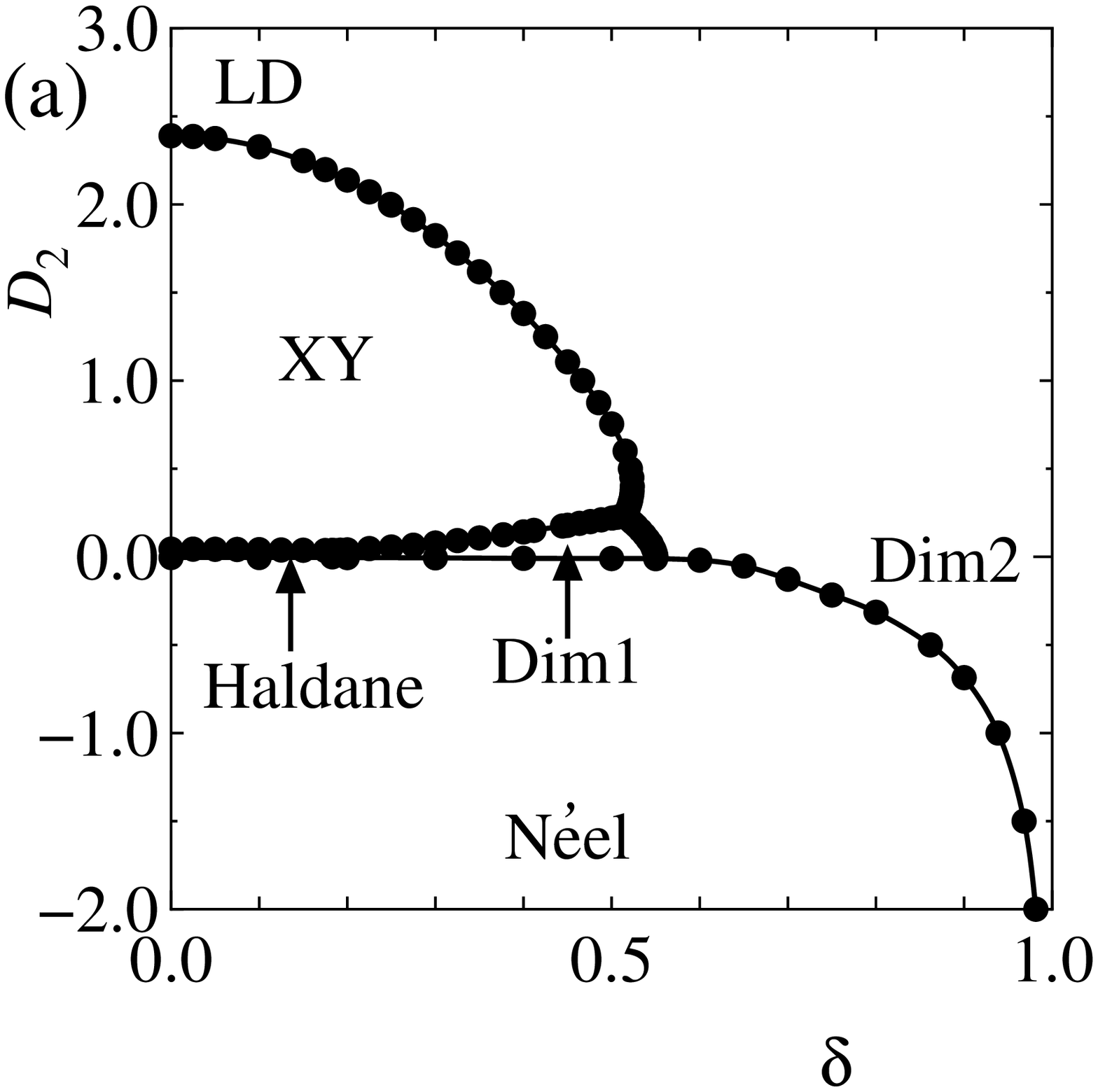}%
       \includegraphics[width=10pc]{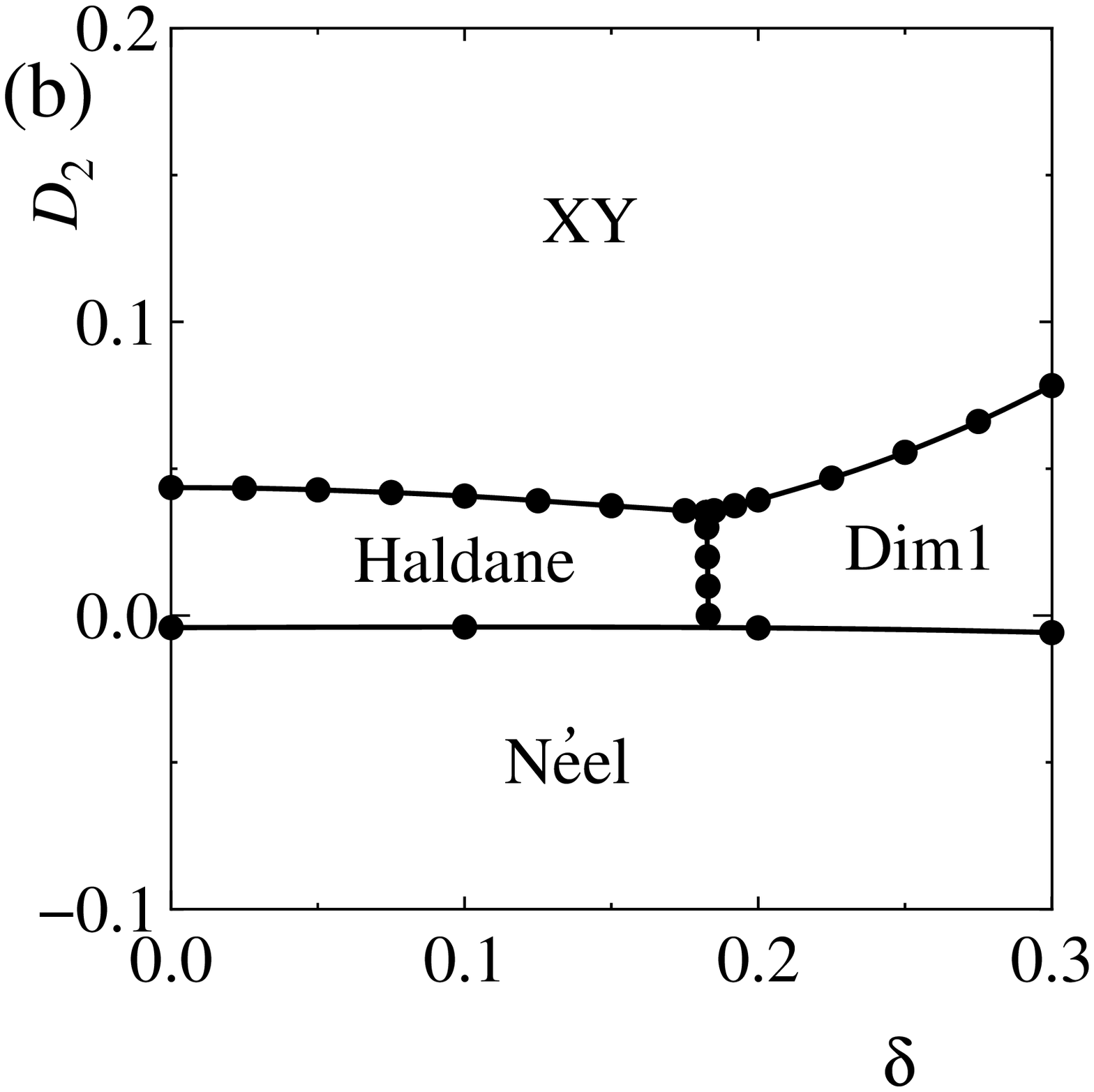}}%
       \caption{
       GS phase diagram on the $\delta-D_2$ plane for $\Delta=1$ and $D_4=0$.
       (a) is for a wide view, while (b) is an enlarged view near the Haldane-Dim1 boundary.
       }
       \label{fig:fig4}
\end{figure}

In the following,
we show the GS phase diagrams of the Hamiltonian (\ref{eq:Ham-3}) for
vairous cases.
Figure \ref{fig:fig4} is the GS phase diagram on the $\delta-D_2$ plane for $\Delta=1$ and $D_4=0$.
In case of $D_2=0$,
the critical values of $\delta$ for the Haldane-Dim1 transition
and the Dim1-Dim2 transition show very good agreements with those of Ref.\citen{yamanaka,yamamoto,kitazawa,nakamura}.
This figure shows that the LD state and the Dim2 state belong to the same phase.
There is no direct path to connect to Haldane region and the LD/Dim2 region on this parameter plane.
Also, the ID phase does not exist on this plane.

\begin{figure}[ht]
       \centerline{
       \includegraphics[width=10pc]{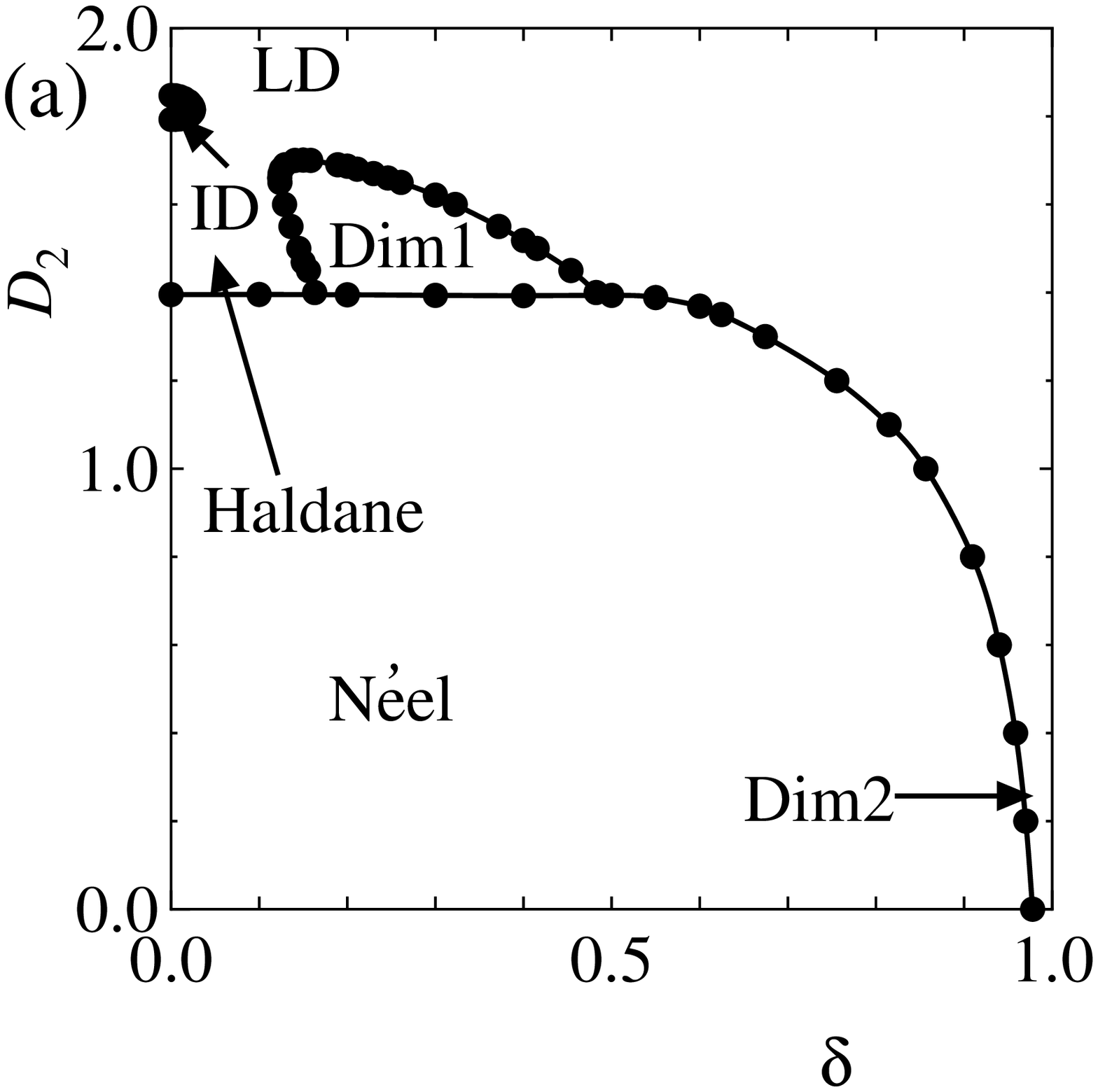}%
       \includegraphics[width=10pc]{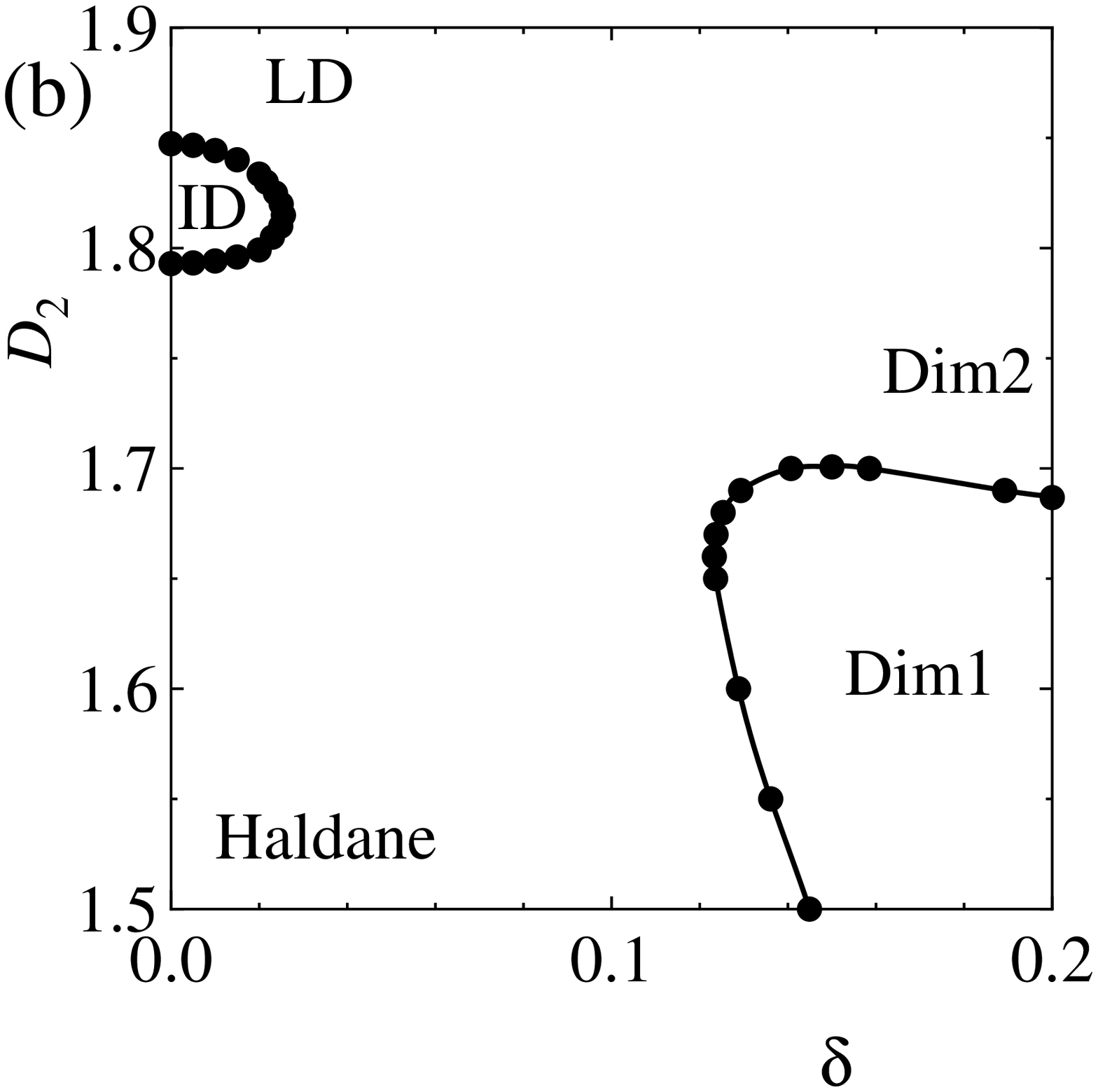}}%
       \caption{
       GS phase diagram on the $\delta-D_2$ plane for $\Delta=2.2$ and $D_4=0$.
       (a) is a wide view, while (b) is an enlarged view for a regime where the trivial states and the SPT states
       strongly compete with each other.
       }
       \label{fig:fig5}
\end{figure}

Figure \ref{fig:fig5} is the GS phase diagram on the $\delta-D_2$ plane for $\Delta=2.2$ and $D_4=0$,
where the ID phase appears.
This GS phase diagram clearly shows that the Haldane state, the LD state and the Dim2 state
belong to the same phase.
Since these three regions are connected to one another,
the direct path connecting the ID region and the Dim1 region cannot exist on this plane.

\begin{figure}[ht]
       \centerline{
       \includegraphics[width=9pc]{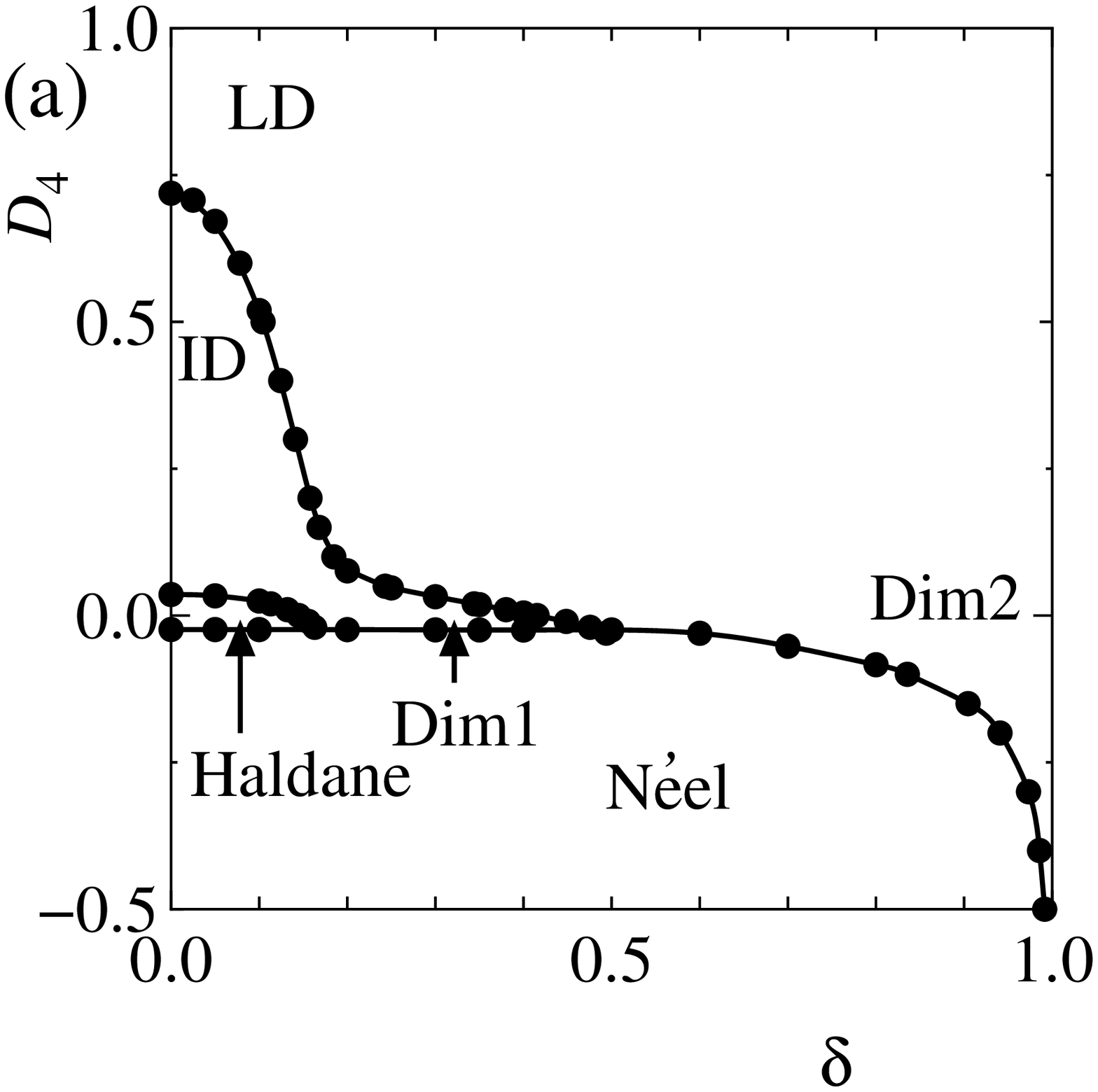}%
       \includegraphics[width=9pc]{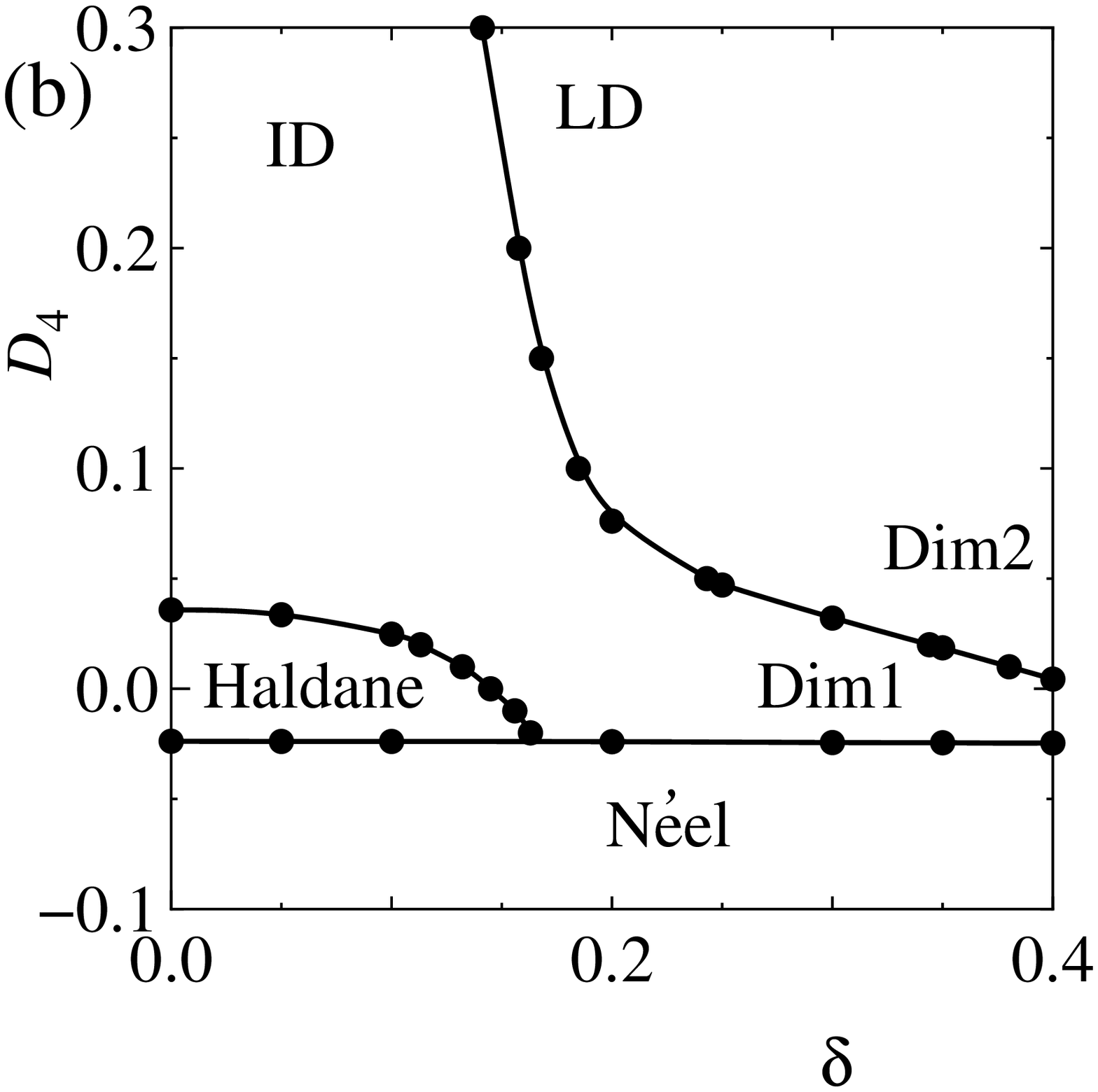}}%
       \caption{
       GS phase diagram on the $\delta-D_4$ plane for $\Delta=2.2$ and $D_2=1.5$.
       (a) is a wide view, while (b) is an enlarged view for a regime where the trivial states and the SPT states
       strongly compete with each other.
       }
       \label{fig:fig6}
\end{figure}

We show the GS phase diagram on the $\delta-D_4$ plane for $\Delta = 2.2$, $D_2 = 1.5$ in Fig.\ref{fig:fig6}.
We see that the ID state and the Dim1 state belong to the same phase.

Let us explain how we determined the phase boundaries.
The quantities $E_0(N,M;{\rm PBC})$ and $E_1(N,M;{\rm PBC})$ represent, respectively,
the lowest and the first excited energies within the subspace of the Hamiltonian $\cH_3$
determined by $N$ and $M$ under the periodic boundary condition (PBC), $\vS_{N+1} = \vS_1$.  
The quantity $N$ is the total number of $S=2$ spins which is supposed to be even,
while $M$ is the total magnetization defined by $M = \sum_j S_j^z$ 
which is a good quantum number.
Similarly we define $E_0(N,M,P;{\rm TBC})$ as the lowest energy 
within the subspace determined by $N$, $M$ and $P$ under the twisted boundary condition (TBC),
$S_{N+1}^x = -S_1^x$, $S_{N+1}^y = -S_1^y$ and $S_{N+1}^z = S_1^z$.
Here $P$ ($P=+1,-1$) is the eigenvalue of the space inversion operator $\cI$
which works as $S_j^\mu \leftrightarrow S_{N-j+1}^\mu$ and commutes with $\cH_3$
under both of the PBC and the TBC. 
We note that this operator $\cI$ is closely related to the condition (iii) of Pollmann {\it et al}.

The quantum phase transitions between the trivial phase and the SPT phase are of the Gaussian type,
and those between the $XY$ phase and one of the above two phases are of the Berezinskii-Kosterlitz-Thouless
(BKT) type.\cite{berezinskii,KT}
In the LS method,\cite{okamoto-ls,nomura-ls,kitazawa-ls,nomura-kitazawa-ls}
we should compare three excitation energies
\begin{eqnarray}
    &&\hskip-1cm\Delta E_{02}^{\rm PBC}(N) \equiv E_0(N,2;{\rm PBC}) - E_0(N,0;{\rm PBC}), \\
    &&\hskip-1cm\Delta E_{00}^{\rm T-P}(N,+1) \equiv E_0(N,0,+1;{\rm TBC}) - E_0(N,0;{\rm PBC}), \\
    &&\hskip-1cm\Delta E_{00}^{\rm T-P}(N,-1) \equiv E_0(N,0,-1;{\rm TBC}) - E_0(N,0;{\rm PBC}),
\end{eqnarray}
in the $N \to \infty$ limit.
Namely, the ground state is one of the XY, trivial, and SPT states
depending on whether $\Delta E_{02}^{\rm PBC}(N)$, $\Delta E_{00}^{\rm T-P}(N,+1)$
or $\Delta E_{00}^{\rm T-P}(N,-1)$ is the lowest among them. 
A physical and intuitive explanation for this method was given in our previous paper.\cite{oka2}
Although we explained by use of the space inversion operator $\cI$ in Ref.\citen{oka2},
a similar explanation is possible by use of the time reversal operator $\cT$ acting as
$\vS_j \to -\vS_j$, which is closely related to the condition (ii) of Pollmann {\it et al}.

\begin{figure}[ht]
       \centerline{
       \includegraphics[height=9pc]{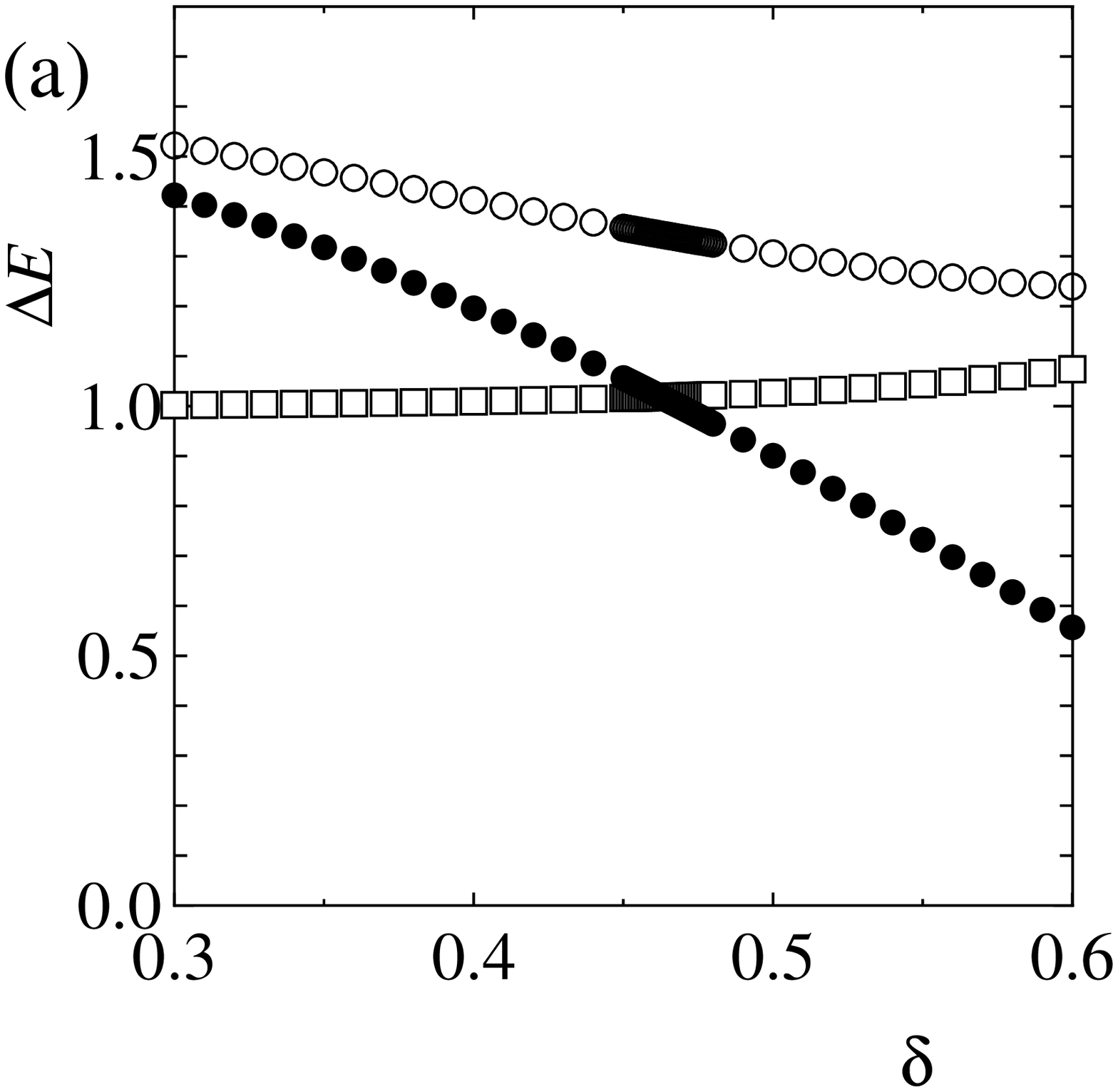}%
       \includegraphics[height=9pc]{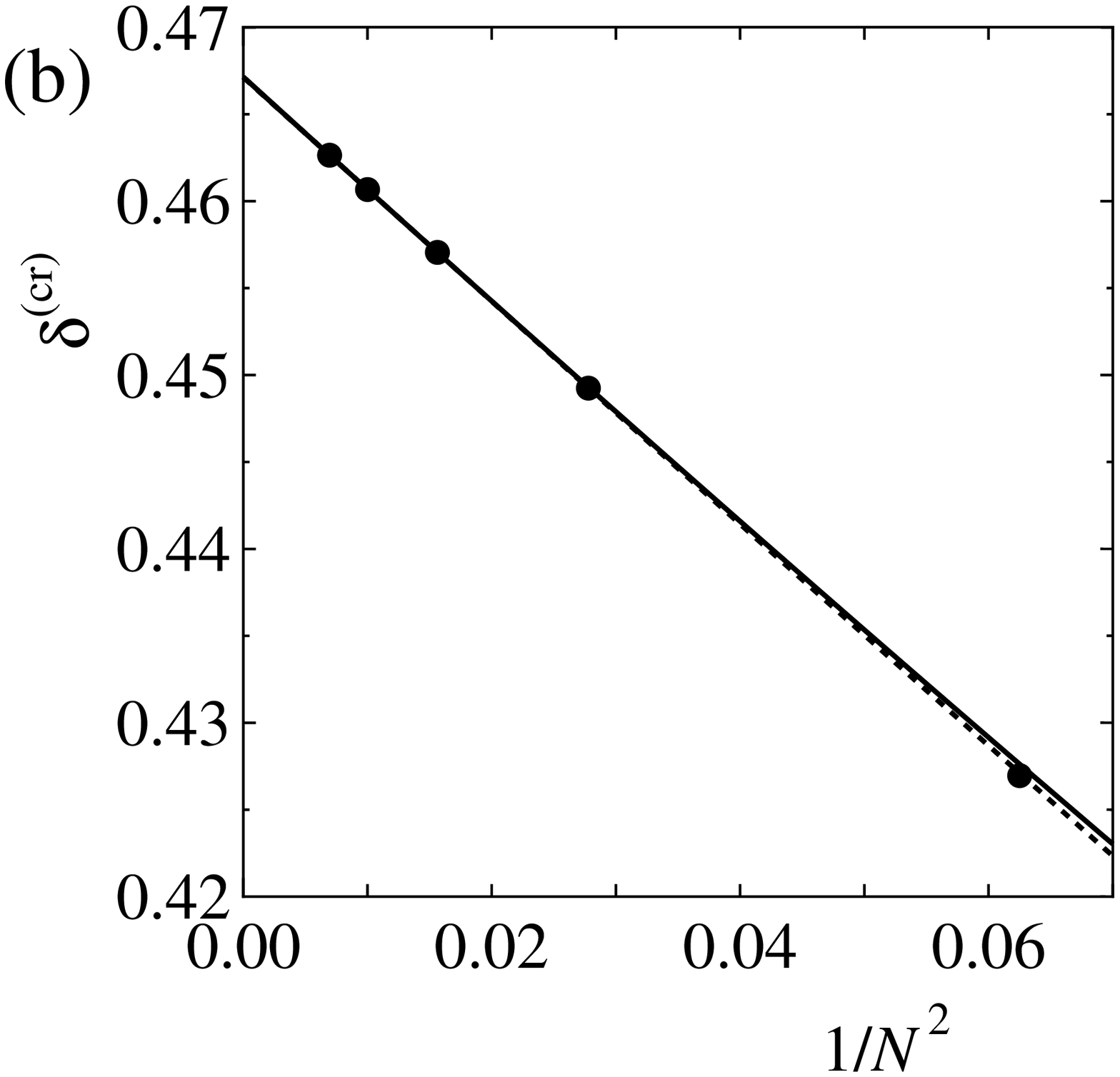}}%
       \caption{An example of the determination of the $XY$-trivial phase transition point in Fig.4,
       where $\Delta = 1$, $D_2=1$ and $D_4=0$.
       In (a), open squares, closed circles and open circles represent, respectively,
       $\Delta E_{02}^{\rm PBC}(N)$, $\Delta E_{00}^{\rm T-P}(N,+1)$ and $\Delta E_{00}^{\rm T-P}(N,-1)$,
       where $N=12$.
       From the crossing point of two curves with open squares and closed circles,
       we obtain $\delta^{\rm(cr)}(N=12) = 0.462647$.
       In (b), we show the extrapolation of $\delta^{\rm (cr)}(N)$ to $N \to \infty$,
       supposing that $\delta^{\rm (cr)}(N)$ is a quadratic function of $1/N^2$.
       The broken line represents the least square result by use of $N=6,8,10$ and 12 data,
       whereas the solid line represent that without $N=6$ data.
       Both lines almost overlap with each other.
       }
       \label{fig:XY-trivial}
\end{figure}

Figure \ref{fig:XY-trivial} shows an example of the determination of
an $XY$-trivial transition point in Fig.\ref{fig:fig4},
where we choose  $\Delta = 1$, $D_2=1$ and $D_4=0$.
From the level crossing between $\Delta E_{02}^{\rm PBC}(N)$
and $\Delta E_{00}^{\rm T-P}(N,+1)$ in Fig.\ref{fig:XY-trivial}(a),
we obtain $\delta^{\rm (cr)}(N)$.
We can estimate the transition point of the infinite system by extrapolating $\delta^{\rm (cr)}(N)$ to $N \to \infty$
as shown in Fig.\ref{fig:XY-trivial}(b),
resulting in $\delta^{\rm (cr)} = 0.4671 \pm 0.0001$

\begin{figure}[ht]
       \centerline{
       \includegraphics[height=9pc]{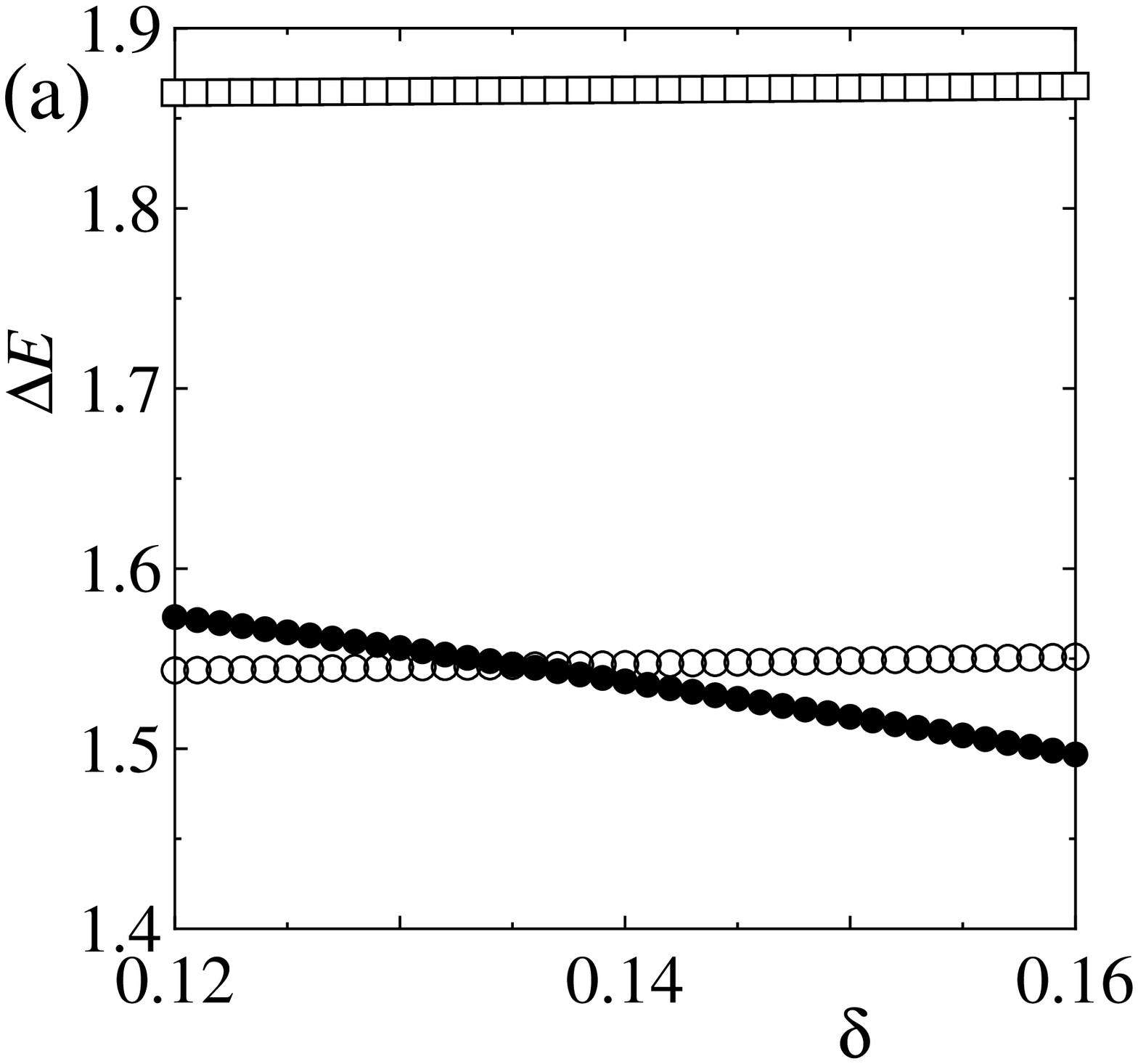}%
       \includegraphics[height=9pc]{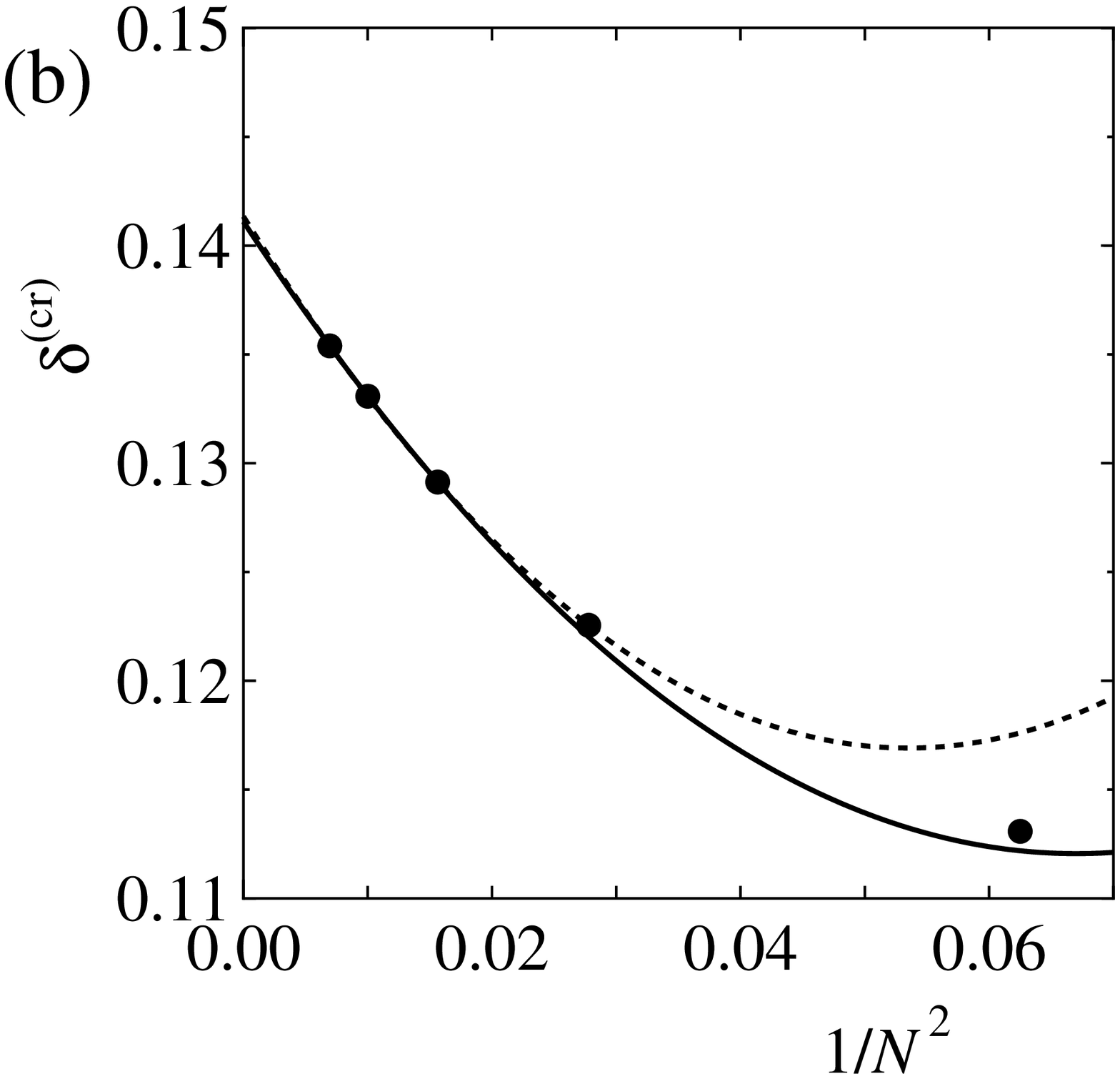}}%
       \caption{An example of the determination of the SPT-trivial phase transition point in Fig.6
       in case of $\Delta = 2.2$, $D_2=1.5$ and $D_4 = 0.3$.
       The meanings of symbols and lines are the same as those in Fig.\ref{fig:XY-trivial}.
       From the crossing point of two curves with open circles and closed circles in (a),
       we obtain $\delta^{\rm(cr)}(N=12) = 0.135395$.
       }
       \label{fig:SPT-trivial}
\end{figure}

Figure \ref{fig:SPT-trivial} shows the method of determining the SPT-trivial
phase transition in Fig.\ref{fig:fig6},
where we choose  $\Delta = 2.2$, $D_2=1.5$.
The level crossing between $\Delta E_{00}^{\rm T-P}(N,+1)$ and
$\Delta E_{00}^{\rm T-P}(N,-1)$ in Fig.\ref{fig:SPT-trivial}(a)
brings about $\delta^{\rm (cr)}(N)$.
The transition point of the infinite system is given by extrapolating $\delta^{\rm (cr)}(N)$ to $N \to \infty$
as shown in Fig.\ref{fig:SPT-trivial}(b),
resulting in $\delta^{\rm (cr)} = 0.1412 \pm 0.0001$.

\begin{figure}[ht]
       \centerline{
       \includegraphics[height=9pc]{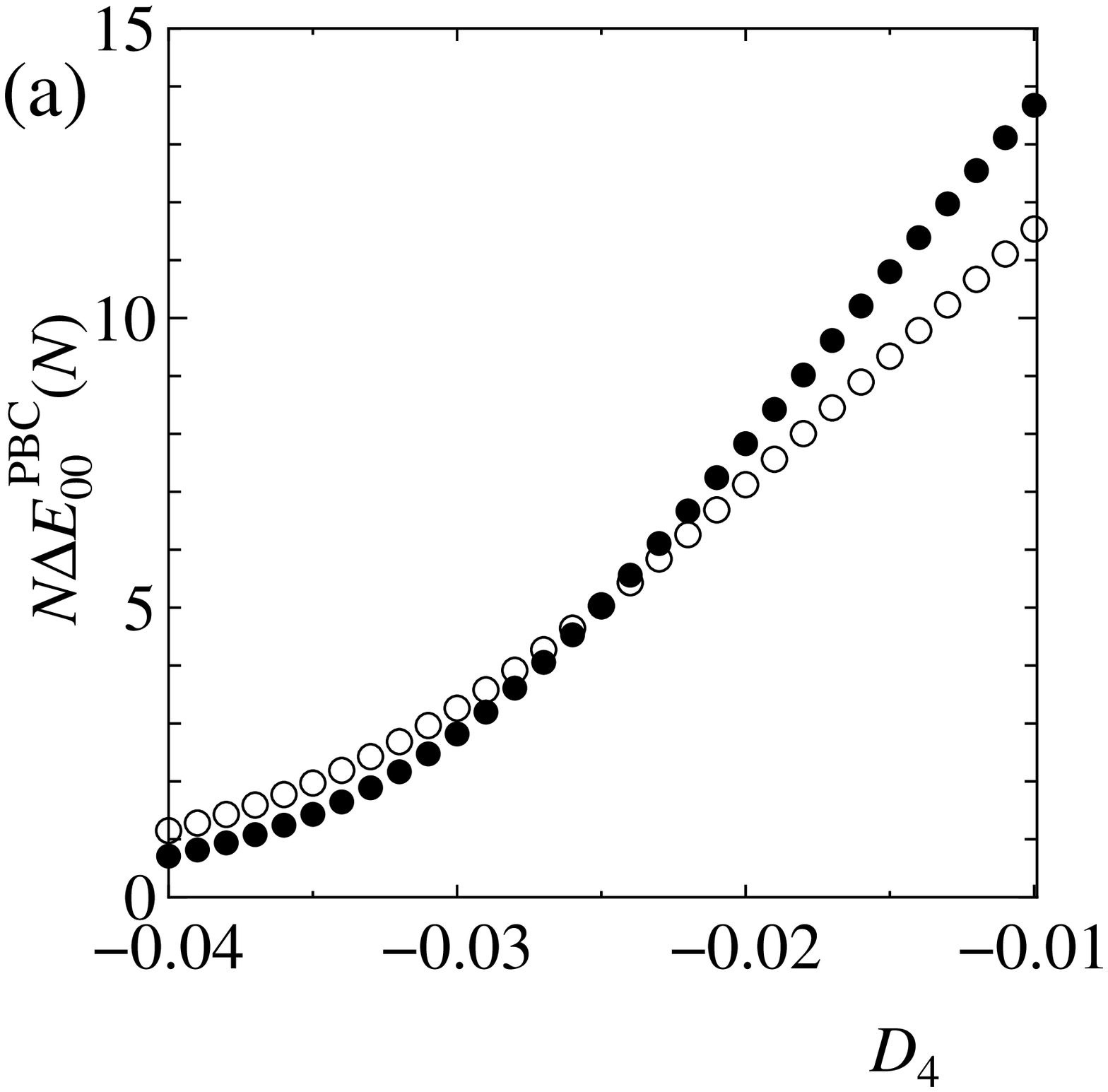}%
       \includegraphics[height=9pc]{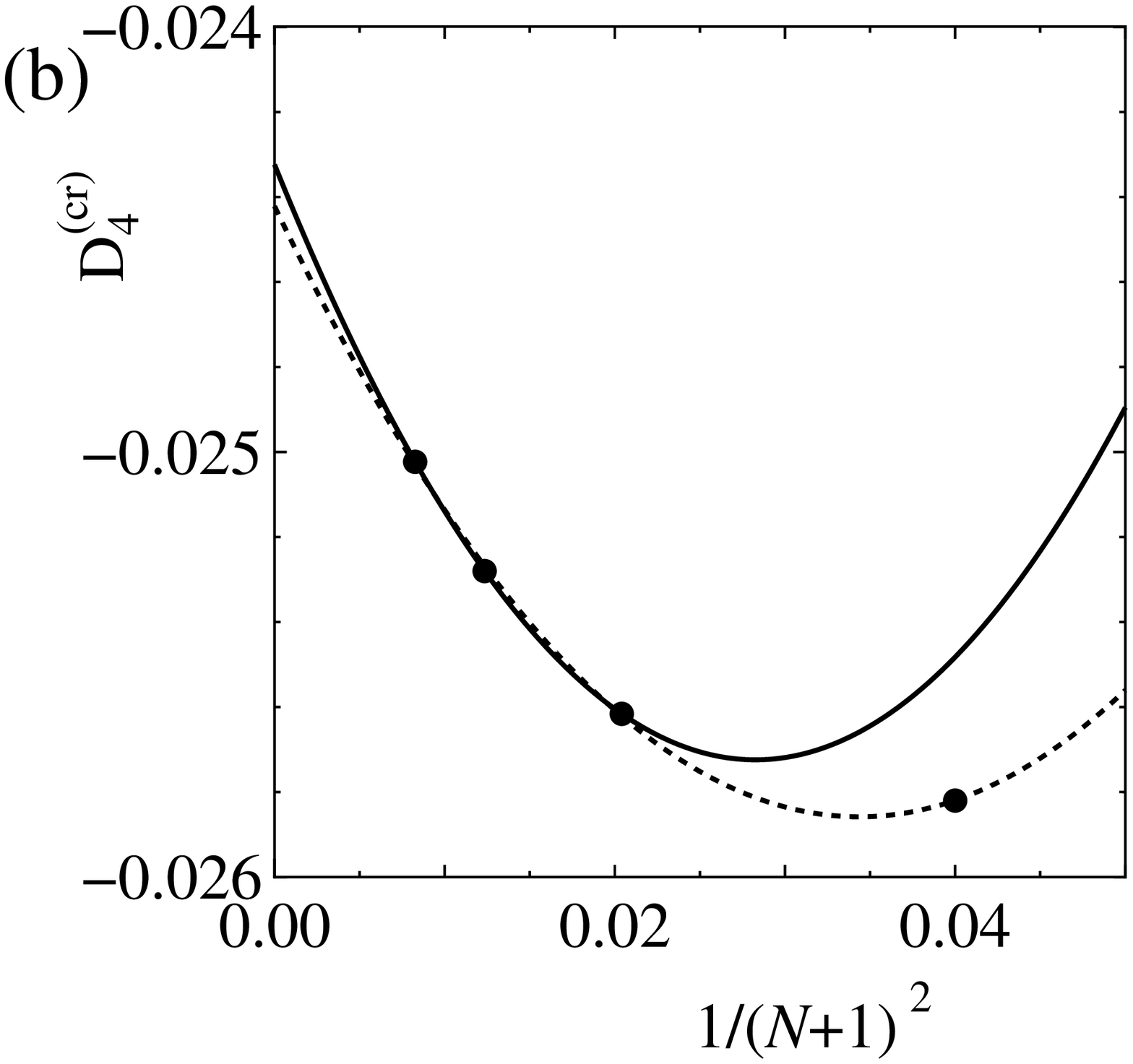}}%
       \caption{An example of the determination of the N\'eel-SPT phase transition point in Fig.\ref{fig:fig6}.
       In (a), open circles and closed circles represent $N=10$ and $N=12$ points, respectively.
       We choose $\Delta = 2.2$, $D_2=1.5$ and $\delta = 0.3$.
       From the crossing point of two curves,
       we obtain $D_4^{\rm (cr)}(N+1=11) = -0.025023$.
       In (b), the broken line represents the least squares result by use of $N+1=5,7,9$ and 11 data,
       while the solid line that without $N+1=5$ data.}
       \label{fig:Neel-SPT}
\end{figure}

Both of the N\'eel-trivial phase transition and the N\'eel-SPT phase transition 
are of the 2D Ising type.
We note that the N\'eel state is a doubly degenerate gapped state,
while the trivial state and the SPT state are unique gapped states.
For determining the 2D Ising phase transition points,
the phenomenological renormalization group (PRG) method\cite{PRG} is useful.
Namely, for instance, running $D_4$ with fixing the parameters $\Delta, \delta$ and $D_2$,
we have numerically solved the PRG equation
\begin{equation}
   N\Delta E_{00}^{\rm PBC}(N) = (N+2)\Delta E_{00}^{\rm PBC}(N+2)  \\
\end{equation} 
to obtain the finite-size critical value $D_4^{\rm cr}(N+1)$ for given $\Delta, \delta$ and $D_2$,
where $E_{00}^{\rm PBC}(N)$ is defined by
\begin{equation}
   \Delta E_{00}^{\rm PBC}(N)
   \equiv E_1(N,0;{\rm PBC}) - E_0(N,0;{\rm PBC}).
\end{equation}
Then, we have estimated the critical value $D_4^{\rm (cr)}$  
by taking the $N \to \infty$ limit,
assuming that the $N$-dependence of $D_4^{\rm (cr)}(N+1)$ is a quadratic function of $1/(N+1)^2$.
Figure \ref{fig:Neel-SPT} shows an example of determining the N\'eel-SPT phase transition point in Fig.6
in case of $\Delta = 2.2$, $D_2=1.5$ and $\delta = 0.3$.
From Fig.\ref{fig:Neel-SPT}(b), we estimate $D_4^{\rm (cr)} = -0.0244 \pm 0.0001$.

\begin{figure}[ht]
       \centerline{
       \includegraphics[width=9pc]{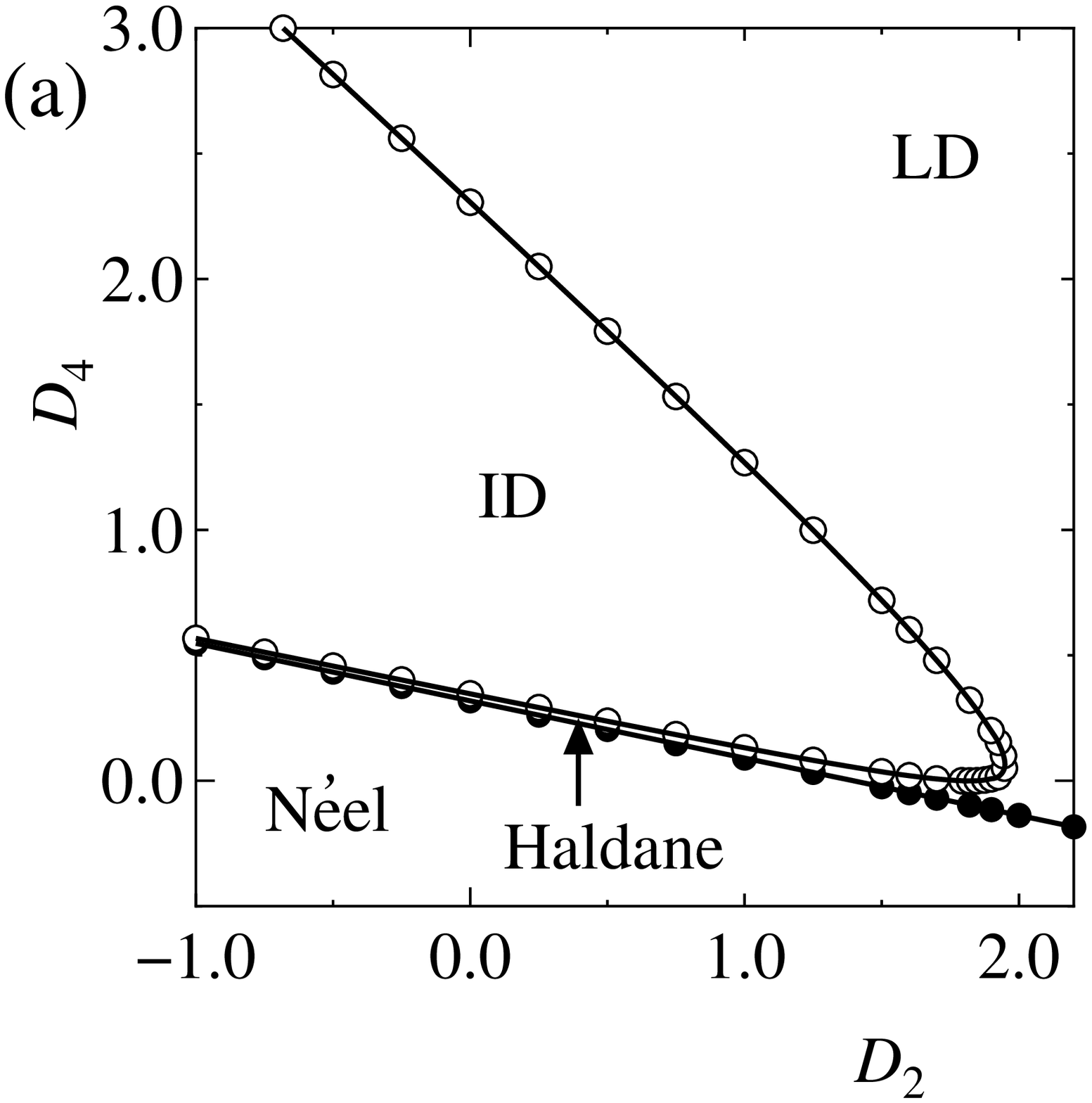}%
       \includegraphics[width=9pc]{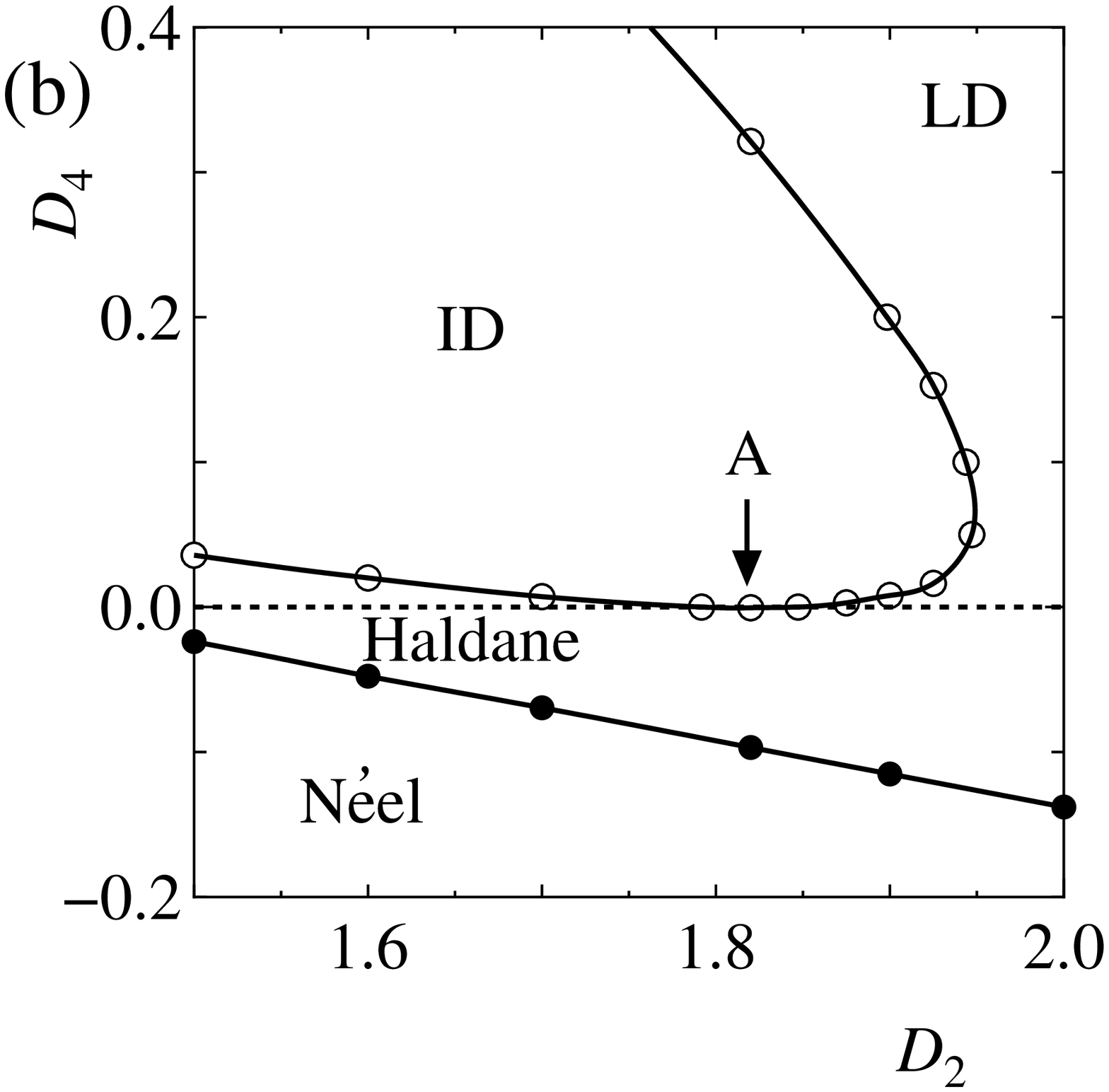}}%
       \caption{GS phase diagram on the $D_2-D_4$ plane in case of $\Delta = 2.2$ and $\delta = 0$.
       (a) is for a wider region, and (b) is an enlarged view near the bottom of the ID region.
       In (b), the point A shows the bottom of the ID region, 
       the $D_4$ coordinate of which is $D_4 = -0.0005 \pm 0.0001$.
       The horizontal broken line shows $D_4=0$.}
       \label{fig:D4-effect}
\end{figure}

Here we explain the reason why we have introduced the $D_4$ term to obtain the GS phase diagram of
Fig.\ref{fig:fig6}, where the ID state and the Dim1 state are directly connected.
Figure \ref{fig:D4-effect} show the GS phase diagram on the $D_2-D_4$ plane in case of
$\Delta = 2.2$ and $\delta = 0$.
For instance, the ID region exists for
$1.7917 < D_2 < 1.8473$
when $D_4=0$,
while it exists for
$0.5188 < D_2 < 1.8984$
when $D_4=0.2$.
We can see that the positive $D_4$ considerably widens the ID region,
as we have already pointed out in Ref.\citen{oka3}.
Thus, the introduction of the $D_4$ term is very crucial to obtain the GS phase diagram
in which the ID state and the Dim1 state are directly connected.
We note that the addition of the $D_4$ term does not change
any symmetry of the Hamiltonian without the $D_4$ term.

Figure \ref{fig:D4-effect} also well explains the difference
between our\cite{tone,oka1,oka2} and Tzeng's\cite{tzeng} conclusion
and Kj\"all {\it et al.}'s\cite{kjall} one with respect to the existence
of the ID phase on the $\Delta-D_2$ plane
in case of $D_4 = \delta =0$, which we already stated.
The $D_4$ coordinate of the point A in Fig.\ref{fig:D4-effect}(b) is $D_4^{\rm (A)} = -0.0005 \pm 0.0001$,
where A is the bottom of the ID region.
We believe that our result is correct, 
but it is possible that a very small upward shift of the point A
by $\Delta D_4^{\rm (A)} = 0.0005$ (or larger) obtained by other methods 
brings about the absence of the ID region on the $\Delta-D_2$ plane with $D_4 = \delta =0$.
Since the fact that $D_4^{\rm (A)}$ is very near to $D_4=0$ seems to be accidental,
we think that the above difference is not a serious problem.

Finally we shortly mention the $S=1$ case.
Tonegawa {\it et al}.\cite{tone2} and  Chen {\it et al}.\cite{chen}
investigated the GS phase diagram of the bond alternating $S=1$ quantum spin chain
with the on-site anisotropy described by
\begin{equation}
    \cH_4
    = \sum_j [1+(-1)^j \delta] \vS_j \cdot \vS_{j+1} + D_2 \sum_j (S_j^z)^2,~~~~S=1.
    \label{eq:Ham-4}
\end{equation}
They obtained the GS phase diagram on the $\delta-D_2$ plane and
showed that the LD state and the dimer state belong to the same phase.
We note that, for the $S=1$ case,
the Haldane state is the SPT state, and the LD state and the dimer state are the trivial states.

In conclusion, 
employing the LS and PRG analysis based on the exact diagonalization
calculation,
we have determined the GS phase diagrams of Hamiltonian (\ref{eq:Ham-3})
on the $\delta-D_2$ plane for the $(\Delta,D_4) = (1,0)$ case (Fig.\ref{fig:fig4})
and the $(\Delta,D_4) = (2.2,0)$ case (Fig.\ref{fig:fig5}),
and on the $\delta-D_4$ plane for the $(\Delta,D_2) = (2.2,1.5)$ case  (Fig.\ref{fig:fig6}).
We have proved that the Haldane state, the LD state and the Dim2 states
belong to the same trivial phase by showing the existence of adiabatic paths directly connecting these
states without the quantum phase transition.
In a similar way,
we have proved that the ID state and the Dim1 state belong to the same SPT phase.

\begin{acknowledgment}
We would like to express our appreciation to Masaki Oshikawa and  Shunsuke C. Furuya
for stimulating discussions.
We thank the Supercomputer Center, Institute for
Solid State Physics, University of Tokyo, and the Computer Room, Yukawa Institute
for Theoretical Physics, Kyoto University, for computational facilities.

\end{acknowledgment}

\end{document}